\documentclass[reprint,
amsmath,amssymb,
aps,
prl,
]{revtex4-1}

\usepackage{graphicx}
\usepackage{dcolumn}
\usepackage{bm}

\usepackage[utf8]{inputenc}
\usepackage[T1]{fontenc}
\usepackage{mathptmx}
\usepackage{amssymb,amsmath}
\usepackage{amsfonts}

\usepackage{easyReview}

\renewcommand{\vec}[1]{\mathbf{#1}}

\begin{document}
	
	
	\title{Inverted Sedimentation of Active Particles in Unbiased ac Fields}
	
	\author{Jos\'e~Carlos~\surname{Ure$\tilde{\text{n}}$a~Marcos}}
	\author{Benno~Liebchen}%
	\email[]{benno.liebchen@pkm.tu-darmstadt.de}
	\affiliation{%
		Institut f\"ur Physik Kondensierter Materie, Technische Universit\"at Darmstadt, 
		Hochschulstr. 8, 64289 Darmstadt, Germany
	}%
	
	\date{\today}
	
	\begin{abstract}
		Achieving control of the motion of active particles is crucial for applications ranging from targeted cargo delivery to nanomedicine.
		While much progress has been made recently to control active motion based on external forces, flows or gradients in concentration or light intensity, which all have a well-defined direction or bias, little is known about how to steer active particles in situations where no permanent bias can be realized. 
		Here, we show that ac fields with a vanishing time average provide an alternative route to steering active particles.
		We exemplify this route for inertial active particles in a gravitational field, observing that a substantial fraction of them 
		persistently travels in the upward direction upon switching on the ac field, resulting in an inverted sedimentation profile at the top wall of a confining container. 
		Our results offer a generic control principle which could be used in the future to steer active motion, to direct collective behaviors and to purify mixtures.
	\end{abstract}
	
	\maketitle
	
	\textit{Introduction.}---Active particles (APs) use energy from their environment to create directed motion. 
	Examples comprise living organisms~\cite{Marchetti2013,Klotsa2019,Nachtigall2001}, both macroscopic -- like the birds which form a flock~\cite{Cavagna2014} -- and microscopic -- such as sperm cells~\cite{Rode2019} and motile bacteria~\cite{Beer2019}, as well as synthetic APs, like motile robots~\cite{mijalkov2016engineering,Scholz2018}, 
	Janus colloids~\cite{Zhang2017,Paxton2004,Howse2007,Bechinger2016,Shelke2019,Liebchen2021,Vutukuri2020} and droplet swimmers~\cite{Izri2014,Cira2015,Kruger2016,Wang2016,Jin2018,Hokmabad2021,Suda2021,Cholakova2021}.
	
	While biological microswimmers can steer autonomously and use this ability to perform sophisticated tasks~\cite{Elgeti2015}, including food search~\cite{Sourjik2012,Cremer2019}, target detection~\cite{Alvarez2014} and the coordination of their collective behavior through communication~\cite{Noorbakhsh2015,Eidi2017}, synthetic APs rely on external control schemes to be able to perform tasks like targeted drug delivery~\cite{Ghosh2020}, microsurgery~\cite{Vyskocil2020} and microplastic collection~\cite{wang2019photocatalytic} in the future. 
	Thus, following the importance of externally steering active motion, a wide range of control schemes has been recently developed. In particular, it is now well known that APs can be controlled via external fields, such as electric fields~\cite{Zhang2022}, rotating magnetic fields~\cite{Palacci2013,Driscoll2017,Martinez-Pedrero2017} or light intensity gradients~\cite{Nedev2015,Zong2015,Moyses2016,Liu2016,Liu2018,Mousavi2019,Buttinoni2022}, as well as with combinations of electric and magnetic fields, which allow them to collect, transport and release cargo at intended locations~\cite{Demirors2018}. 
	There are now also established ways to steer APs by topographical features in their environment~\cite{Simmchen2016,Wu2018,Wang2021}, boundaries~\cite{das2015boundaries}, and feedback-control systems~\cite{Khadka2018,fernandez2020feedback,lavergne2019group}, as well as with external stimuli (gradient fields) acting on the swimming direction of synthetic APs via tactic phenomena, including chemotaxis~\cite{hong2007chemotaxis,liebchen2018synthetic,moeller2021}, phototaxis~\cite{Lozano2016phototaxis,GomezSolano2017,Lozano2019,Jahanshahi2020}, gravitaxis~\cite{TenHagen2014,ruehle2022}, thermotaxis~\cite{auschra2021thermotaxis} or viscotaxis~\cite{Liebchen2018,Datt2019,stehnach2021viscophobic}.
	
	Notably, all these control schemes involve fields with a well-defined direction or bias, which may be stationary or vary slowly with time (in such a way that APs can follow them adiabatically). Conversely, fast and unbiased ac fields have so far been mainly used to endow particles with the ability to self-propel~\cite{Chen2014,Shields2017,Mano2017,Nadal2020,Wu2020,Lee2021,Nishiguchi2018} or to collect cargo~\cite{Demirors2018}, but hardly to control the dynamics of APs.
	
\begin{figure*}[t]
		\centering
		\includegraphics[width=.9\textwidth]{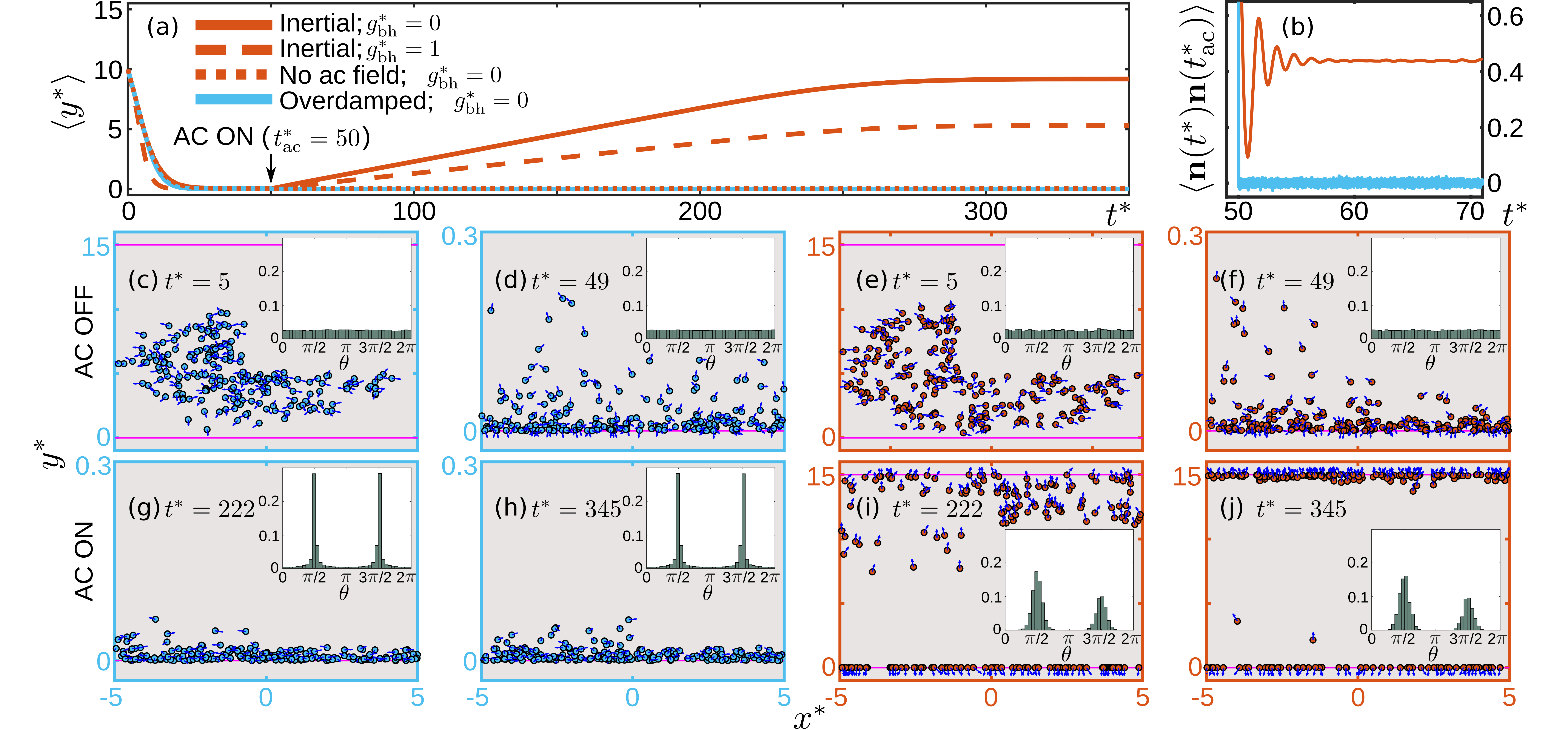}
		\caption{(a) Average vertical position $\langle y^* \rangle$ of $10^4$ inertial (red lines and frames) and rotationally overdamped (blue) ABPs. (b) Autocorrelation of the self-propulsion direction after switching on the ac field. (c-j) Simulation snapshots of representative subsets of 200 ABPs with $g^*_\text{bh}=0$. Dark-blue arrows indicate self-propulsion direction. Insets: distribution of the self-propulsion directions over 20 oscillations of the ac field. Parameter values: $v^*_\text{s}=0.9$, $D^*=0.01$, ${D^*_\text{r}}=1$, ${I^*}^2/(2{\omega^*}^2)=0$ $(t^*<50)$, ${I^*}^2/(2{\omega^*}^2)=20$ $(t^*\geq50)$, ${\omega}^*=500$, timestep $\Delta t^*=10^{-4}/\omega^*$.}
		\label{fig1}
	\end{figure*}

	Here, we show that rapidly oscillating ac fields provide a novel route to controlling the self-propulsion direction of APs without requiring any large-scale gradients, directed flows, forces or torques. Instead, the control principle which we propose hinges on the stabilization of fixed points in the orientation dynamics of APs which would be unstable in the absence of ac fields. 
	To exemplify this control principle, we consider APs sedimenting at the lower wall of a container. When switching on a rapidly oscillating ac field which couples to the orientation of the APs, we observe that most of the particles stop sedimenting and persistently self-propel in the upward direction, resulting in an inverted sedimentation profile at the top wall [Movie S1 and Fig.~\ref{fig1}]. This is achieved by exploiting (weak) inertial effects in APs to stabilize the fixed point corresponding to upward motion. In contrast to previous works~\cite{Wolff2013,TenHagen2014}, our control principle does not require gravitaxis (but is robust against both positive and negative gravitaxis), bottom-heaviness or non-spherical particle shapes, and can also be used e.g. to purify particle mixtures or, as we show, to revert transport.
	
	\textit{Model.}---To exemplify the idea of using ac fields to control self-propulsion, let us consider (inertial) active Brownian particles (ABPs) in two dimensions~\cite{TenHagen2011,Bechinger2016,Lowen2020,Hecht2021}, self-propelling at a constant speed $v_0$ along the direction $\vec{n}(t)=(\cos{\theta(t)},\sin{\theta(t)})$, with $\theta$ being the orientation angle of the particle with respect to the x-axis. The particles are subjected to an external force $\vec{F}$ and a torque $T$, yielding the following equations of motion
	\begin{equation}\label{trans}
		m\ddot{\vec{r}} + \gamma\dot{\vec{r}} = \gamma v_0\vec{n} + \vec{F} + \gamma\sqrt{2D}\boldsymbol{\eta},
	\end{equation}
	\begin{equation}\label{angle}
		J\ddot{\theta} + \gamma_\text{r}\dot{\theta} = T + \gamma_\text{r}\sqrt{2D_\text{r}}{\eta_\text{r}},
	\end{equation}
	where $m$ and $J$ are the mass and the moment of inertia of the particles; $\gamma$, $\gamma_\text{r}$, $D$ and $D_\text{r}$ are, respectively, the translational and rotational damping (Stokes' drag) and diffusion coefficients; and $\boldsymbol{\eta}(t)$ and $\eta_\text{r}(t)$ represent zero-mean, unit-variance Gaussian white noise. This model could be realized e.g. based on autophoretic Janus colloids in a liquid, with light-powered Janus colloids in a gas, or with vibrated granulates. 
	
	\textit{Inverting sedimentation with ac fields.}---As a first example, we consider particles in a gravitational field $\vec{F}=-mg\hat{e}_{y}$, where $g$ is the effective gravitational constant, experiencing a torque $T=T(\theta,t)=I\cos{\theta}\sin(\omega t)-g_\text{bh}\cos{\theta}$, where the first term represents an ac field with frequency $\omega$, strength $I$ and vanishing time average $\langle I\cos\theta\sin(\omega t)\rangle_t=0$, which could be realized e.g. with magnetic colloids~\cite{Palacci2013Living,Han2021,Li2022}, magnetotactic bacteria~\cite{Pierce2017,Petroff2022} or metallodielectric colloids~\cite{Behdani2021} in time-dependent magnetic or electric fields respectively, or with magnetized granular particles in ac magnetic fields~\cite{ledesma2022magnetized} on (tilted) vibrating plates~\cite{Scholz2018,Scholz2018a}, as we further specify below. The second term represents an optional (downward) bias which can take a non-zero value e.g. for bottom-heavy Janus particles~\cite{Campbell2013,Singh2018,Rashidi2020} or shape-asymmetric vibrated granulates~\cite{TenHagen2014,Scholz2018,Scholz2018a} on a tilted plate.
	
	To reduce the parameter space, we now rescale space and time as $t^*=\gamma_\text{r}t/J$ and $\vec{r}^*
	=\gamma_\text{r}\vec{r}/(v_0J)$, which simplifies Eqs.~(\ref{trans}) and~(\ref{angle}) to (note that $\boldsymbol{\eta}(t)=\sqrt{\gamma_\text{r}/J}\boldsymbol{\eta}(t^*)$):
	$    m^*\ddot{\vec{r}}^* + \dot{\vec{r}}^* = \vec{n} - v^*_\text{s}\hat{e}_{y} + \sqrt{2D^*}\boldsymbol{\eta}$,
	$    \ddot{\theta}+\dot{\theta}=\big[-g^*_\text{bh} + I^*\sin{(\omega^*t^*)}\big]\cos{\theta} + \sqrt{2D^*_\text{r}}{\eta_\text{r}},
	$
	where overdots denote the derivative with respect to $t^*$. The dimensionless parameters $m^*=\gamma_\text{r}m/(J\gamma)$, $v^*_\text{s}=mg/(\gamma v_0)$, $D^* = \gamma_\text{r}D/(Jv^2_0)$, $g^*_\text{bh}=Jg_\text{bh}/\gamma^2_\text{r}$ and $D^*_\text{r} = J{D}_\text{r}/\gamma_\text{r}$ are fixed by the specific system under consideration (see below for typical values), whereas $I^*=JI/\gamma^2_\text{r}$ and $\omega^*=J\omega/\gamma_\text{r}$ can be adjusted via the ac field, thus serving as our key control parameters. In what follows, we neglect translational inertia ($m^*\ddot{\vec{r}}^*=0$), which is unimportant for our results.
	
	We now perform Brownian dynamics simulations of ABPs, with and without rotational inertia, initialized at $\vec{r}^*_0=(x^*_0,y^*_0)=(0,10)$ with uniformly distributed random orientations and vanishing velocities and accelerations. We confine the particles between two horizontal walls placed at $y^*_\text{b}=0$ and $y^*_\text{t}=15$ by setting the vertical component of the particle velocity to zero if a particle moves towards either wall. The ac field is initially off ($I^*=0$), and so the particles move downwards due to the gravitational field [Figs.~\ref{fig1}(a,c,e) and Movie S1] both in the overdamped case ($J=0$) and in the presence of inertia ($J>0$). Once the ABPs have (almost) reached the stationary sedimentation profile [Figs.~\ref{fig1}(d,f)] and their average position is close to the botton wall [$\langle y^* \rangle\approx 0$, Fig.~\ref{fig1}(a)], we switch on the ac field (at $t^*_\text{ac}=50$). While overdamped particles simply continue sedimenting [Figs.~\ref{fig1}(g,h)], strikingly, in the presence of inertia, $\langle y^* \rangle$ suddenly starts increasing [Fig.~\ref{fig1}(a)], until a plateau is reached at $\langle y^* \rangle>y^*_\text{t}/2$. That is, most of the inertial ABPs start to persistently move upwards, against the acting net force, once the ac field is on [Fig.~\ref{fig1}(i)]. This continues until they reach the top wall, resulting in an inverted sedimentation profile coexisting with a remaining (smaller) sedimentation profile at the bottom wall [Fig.~\ref{fig1}(j)]. Remarkably, both profiles are exponential~\cite{Palacci2010,Enculescu2011,Wolff2013,Scagliarini2022}, with the one at the top wall showing a sedimentation length almost two orders of magnitude greater than that at the bottom wall for the chosen parameters (see SM~\cite{[{See Supplemental Material at }][{ for further details of inverted sedimentation, a detailed derivation of Eqs. (\ref{bar_angle_nd}) and (\ref{effective_potential_1st}) and a detailed discussion of inverted transport.}]SM}). Note that the plateau value of $\langle y^* \rangle$ depends not only on the choice of parameters but also on the initial sign of $I^*$ (see SM~\cite{SM}). However, at late times, fluctuations induce random flips between upward and downward motion [see Movie S1], which leads to a slow decay of $\langle y^* \rangle$ to $y^*_\text{t}/2$ after $t^*\sim 10^4$ (not shown).
	
	\textit{ac fields induce persistent orientation}---To understand the observed sedimentation reversal, let us now analyze the distribution of particle orientations. Before switching on the ac field, this distribution is uniform [insets in Figs.~\ref{fig1}(c-f)]. Upon switching on the field, overdamped particles essentially follow the (rapid) oscillations of the ac field, so that they self-propel about half of the time upwards and downwards [insets in Figs.~\ref{fig1}(g,h)]. Thus, their net motion is essentially determined by the competition between the gravitational drift and translational diffusion, so the ac field hardly impacts the sedimentation profile [Figs.~\ref{fig1}(g,h)]. In stark contrast, for inertial APs, the ac field not only creates a significant upward bias [insets in Figs.~\ref{fig1}(i,j)], but also stabilizes the particle orientation, as shown by its autocorrelation function in Fig.~\ref{fig1}(b). This allows them to persistently self-propel towards the top wall [see Movie S1].
	
	\textit{Why do ac fields revert the sedimentation profile?}---To understand the bias and the persistence observed in the particle orientations, we now decompose $\theta$ into a `fast' component $\theta_\text{ac}$ which is expected to oscillate on the same timescale as the ac field, and a `slow' part $\bar{\theta}$ representing the net dynamics after averaging over the ac field; i.e. $\theta=\bar{\theta}+\theta_\text{ac}$.
	Plugging $\theta=\bar{\theta}+\theta_\text{ac}$ into the equations of motion and averaging over the period of the ac field, we obtain (detailed derivation in SM~\cite{SM})
	\begin{equation}\label{bar_angle_nd}
		\ddot{\bar{\theta}}+\dot{\bar{\theta}}=-\partial_{\bar{\theta}} U^*_{\text{eff}}(\bar{\theta}) + \sqrt{2{D^*_\text{r}}}{\eta_\text{r}},
	\end{equation}
	\begin{equation} \label{effective_potential_1st}
		U^*_{\text{eff}}(\bar{\theta})=g^*_\text{bh}\sin{\bar{\theta}}+\frac{{I^*}^2}{4{\omega^*}^2}\cos^2{\bar{\theta}}.
	\end{equation}
	Note that it is sufficient to explore the orientation dynamics of the particles, which is not influenced by their spatial dynamics.
	
\begin{figure}[!h]
		\includegraphics[width=.4\textwidth]{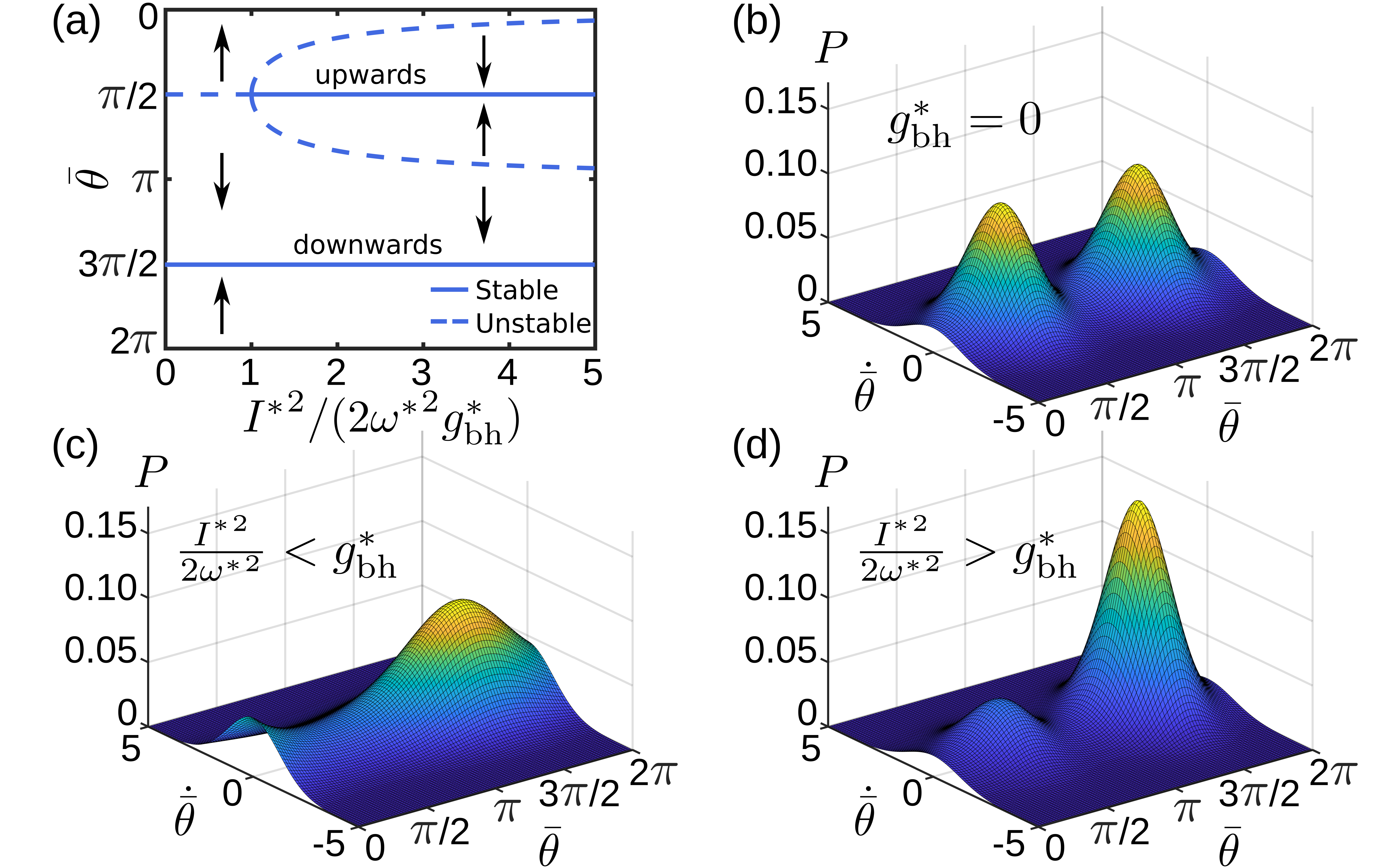}
		\caption{(a) Bifurcation diagram corresponding to Eq. (\ref{bar_angle_nd}) with $D^*_\text{r}=0$. Arrows indicate the flow of $\bar{\theta}$. The unstable fixed points for ${I^*}^2/(2{\omega^*}^2)>g^*_\text{bh}$ are given by $\bar{\theta}=-\arcsin(\frac{2{\omega^*}^2g^*_\text{bh}}{{I^*}^2})+\pi$ and $\bar{\theta}=\arcsin(\frac{2{\omega^*}^2g^*_\text{bh}}{{I^*}^2})$. (b-d) $P(\bar{\theta},\dot{\bar{\theta}})$ for ${D^*_\text{r}}=1.5$ and $g^*_\text{bh}=0$, ${I^*}^2/(2{\omega^*}^2)=5$ (b);
			$g^*_\text{bh}=1$, ${I^*}^2/(2{\omega^*}^2)=0.1$ (c); $g^*_\text{bh}=1$, ${I^*}^2/(2{\omega^*}^2)=5$ (d).}
		\label{fig2}
	\end{figure}

	It is instructive to first discuss the case of vanishing noise (${D^*_\text{r}}=0$). Eqs. (\ref{bar_angle_nd}) and (\ref{effective_potential_1st}) show that $\bar{\theta}_\text{up}=\pi/2$ and $\bar{\theta}_\text{down}=3\pi/2$ are always fixed points. Performing a linear stability analysis of Eq. (\ref{bar_angle_nd}) shows that $\bar{\theta}_\text{down}$ is always stable, whereas $\bar{\theta}_\text{up}$ is stable only if ${I^*}^2/(2{\omega^*}^2)>g^*_\text{bh}$, a result resembling dynamical stabilization ~\cite{Landau1960,Stephenson1908,Simons2009,Butikov2011,Morzuch2012}. This condition leads to a subcritical pitchfork bifurcation, with both fixed points being stable and their basins of attraction equally large for $g^*_\text{bh}=0$ and $|I^*|>0$ [Fig.~\ref{fig2}(a)].
	
\begin{figure}[ht]
		\includegraphics[width=.4\textwidth]{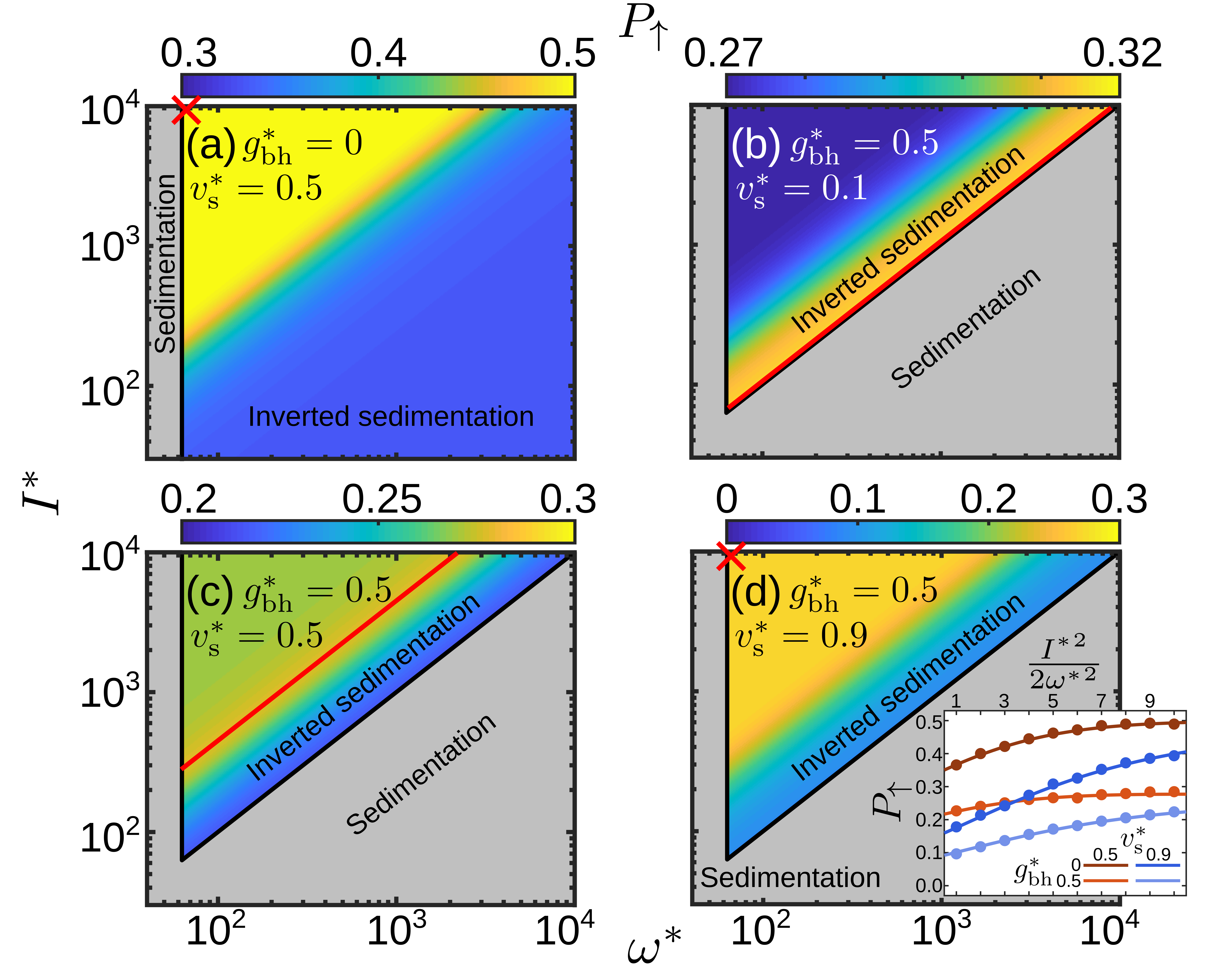}
		\caption{Colors represent $P_\uparrow$ as obtained from Eqs. (\ref{maximum1}) and (\ref{up}). Regions where ${I^*}^2/{2\omega^*}^2<g^*_\text{bh}$ or $\omega^*<10\max(\sqrt{g^*_\text{bh}},2\pi)$ are shown in gray. The maximum of $P_\uparrow$ is marked in red. Inset: $P_\uparrow$ as a function of ${I^*}^2/(2{\omega^*}^2)$ as obtained from Eqs. (\ref{maximum1}) and (\ref{up}) (solid lines) and simulations of $10^4$ ABPs (dots), where we calculate $P_\uparrow$ as the fraction of particles with $\sin(\theta)>v^*_\text{s}$ at the end of the simulation ($t^*=400$). Parameters: $D^*=0.01$, ${D^*_\text{r}}=1$, ${\omega}^*=500$, simulation timestep $\Delta t^*=10^{-2}/\omega^*$, $t^*_\text{ac}=50$.}
		\label{fig3}
	\end{figure}

	To understand the role of noise, we now consider the Fokker-Planck equation equivalent to Eq. (\ref{bar_angle_nd})~\cite{Risken1996}
	\begin{equation}\label{FP_Ueff}
		\frac{\partial P}{\partial {t^*}}=-\dot{\bar{\theta}}\frac{\partial P}{\partial \bar{\theta}}+\frac{\partial}{\partial \dot{\bar{\theta}}}\bigg[\big(\dot{\bar{\theta}}+\frac{dU^*_{\text{eff}}}{d\bar{\theta}}\big)P\bigg]+{D^*_\text{r}}\frac{\partial^2P}{\partial \dot{\bar{\theta}}^2},
	\end{equation}
	which is solved exactly in the steady state with the Ansatz ${P}(\bar{\theta},\dot{\bar{\theta}})=N\exp\big(-\alpha\big[\dot{\bar{\theta}}^2+2U^*_{\text{eff}}(\bar{\theta})\big]\big)$~\cite{Simons2009}. Here, $\alpha=(2D^*_\text{r})^{-1}$ and $N=\frac{1}{\sqrt{2\pi {D^*_\text{r}}}}\big(\int_{0}^{2\pi} \exp\big[-\frac{1}{{D^*_\text{r}}}U^*_{\text{eff}}(\bar{\theta})\big] d\bar{\theta}\big)^{-1}$. Integration over $\dot{\bar{\theta}}$ from $-\infty$ to $\infty$ yields the overall steady-state probability distribution for $\bar{\theta}$
	
	\begin{equation}\label{maximum1}
		P(\bar{\theta})=N_{\bar{\theta}}\exp\big[-\frac{1}{{D^*_\text{r}}}U^*_{\text{eff}}(\bar{\theta})\big],
	\end{equation}
	where $N_{\bar{\theta}}=\sqrt{2\pi {D^*_\text{r}}}N$, which is essentially a Boltzmann distribution for the slow variable $\bar{\theta}$ in the effective potential $U_\text{eff}$. The maxima of ${P}(\bar{\theta},\dot{\bar{\theta}})$ and Eq. (\ref{maximum1}) coincide with the stable equilibrium points encountered for the noiseless case, and we recover the condition ${I^*}^2/(2{\omega^*}^2)>g^*_\text{bh}$ for stabilizing the upper fixed point, as illustrated in Figs.~\ref{fig2}(b-d). Importantly, however, rotational diffusion strongly affects the relative probability of the maxima as it broadens the peaks in Fig.~\ref{fig2} and tends to even them out. This analytical result sheds light on the phenomenon of inverted sedimentation: if ${I^*}^2/(2{\omega^*}^2)>g^*_\text{bh}$ (in physical units $I^2>2J\omega^2g_\text{bh}$) is fulfilled, and $\omega^*\gg\max(\sqrt{g^*_\text{bh}}$,$2\pi)$ (or $J\omega \gg \max(\sqrt{Jg_\text{bh}}$,$2\pi\gamma_\text{r})$) for Eqs. (\ref{bar_angle_nd}) and (\ref{effective_potential_1st}) to be meaningful [see Fig.~\ref{fig3}], the ac field stabilizes upward self-propulsion, and thus the ABPs which happen to be pointing upwards the moment the ac field is switched on persistently travel towards the top wall, in competition with fluctuations.
	
	\textit{Delay effects.}---Let us now provide an explanation of the physical mechanism allowing the ac field to stabilize the upper fixed point. Whereas the orientation of overdamped ABPs essentially follows the oscillations of the ac field, inertial ABPs show a delay in their response to the field~\cite{Caprini2022} [Fig.~\ref{fig1}(b)]. Thus, during each oscillation, there are time instants when the particle orientation turns away from the vertical direction, while the ac field pushes it towards the fixed point. The reverse process, where the particle orientation turns into the vertical position while the ac field pushes it away from the fixed point, is less efficient, since the ac field is weaker close to the fixed point. That is, (tiny) delay effects are crucial to observe inverted sedimentation.
	
	\textit{What controls the fraction of particles moving upwards?}---Particles move upwards if the vertical component of their self-propulsion velocity exceeds the gravitational drift, i.e. if $v_0\sin{\theta}>mg/\gamma$, which leads to a steady-state probability
	\begin{equation}\label{up}
		P_\uparrow=P(\sin(\bar\theta)>v^*_\text{s})=\int_{\arcsin(v^*_\text{s})}^{\pi-\arcsin(v^*_\text{s})}P(\bar{\theta}) d\bar{\theta}.
	\end{equation}
	By numerically evaluating this integral, we can predict the full parameter dependence of the fraction of particles sedimenting at the top [Fig.~\ref{fig3}] in close quantitative agreement with our simulations [inset in Fig.~\ref{fig3}]. Note that, for the comparison with the analytical results, we have averaged the values of $P_\uparrow$ obtained in the simulations over the $I^*<0$ and $I^*>0$ cases to achieve faster convergence of our results (see SM~\cite{SM}). As expected, $P_\uparrow$ broadly decreases with increasing bottom-heaviness ($g^*_\text{bh}$) and sedimentation speed ($v^*_\text{s}$). Strikingly, however, the dependence of $P_\uparrow$ on $I^*$ and $\omega^*$ is highly nontrivial and nonmonotonic, as can be seen in Fig.~\ref{fig3}, where red lines and crosses show the maximum of $P_\uparrow$. Importantly, inverted sedimentation typically sets in with a finite probability $P_\uparrow$ (as typical for subcritical transitions) exceeding 20$\%$ in most cases.
	
\begin{figure}[ht]
		\centering
		\includegraphics[width=0.3\textwidth]{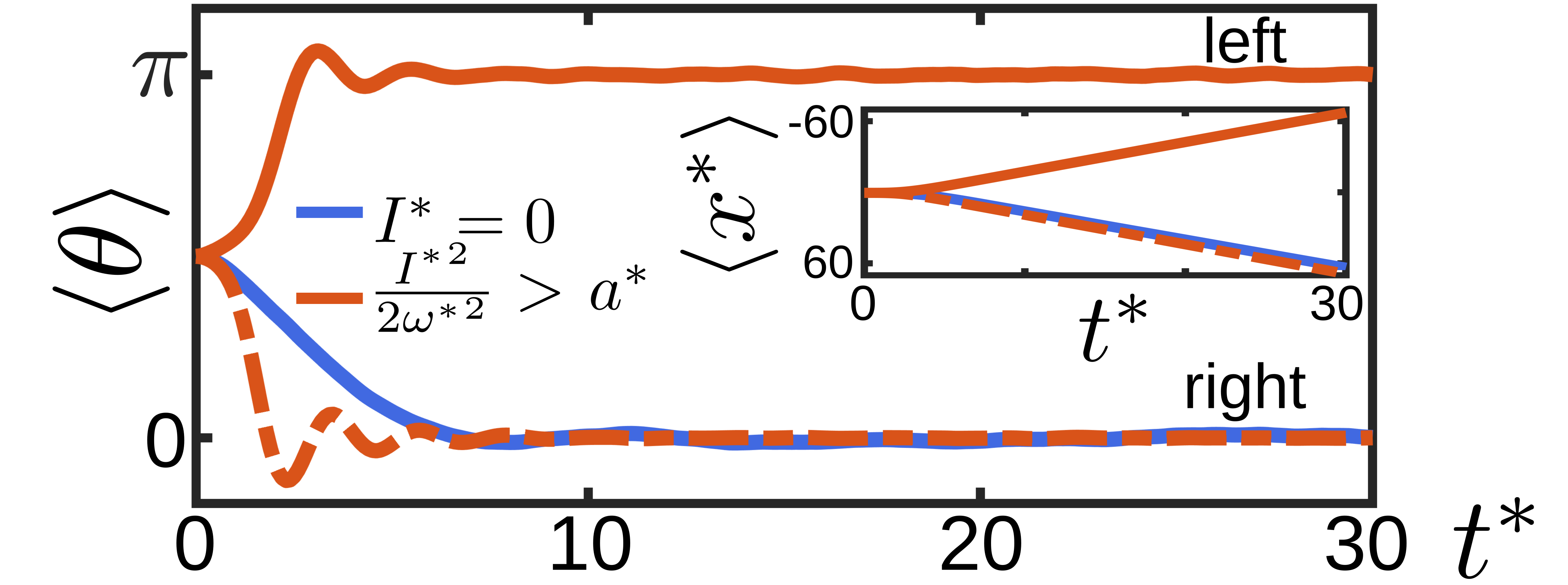}
		\caption{Transport (blue lines and red dashed lines) and inverted transport (red solid lines) of ABPs. Main plot: mean orientation of particles moving to the right and to the left respectively; inset: average trajectories. Details in SM~\cite{SM}.}
		\label{fig4}
	\end{figure}

	\textit{Inverted transport.}---To show the generality of our control principle, in the SM~\cite{SM} we apply rapidly oscillating ac fields to revert the direction of motion of APs in a periodic light intensity field. For $I^*=0$, particles self-propel to the right [blue lines in Fig.~\ref{fig4}], whereas for sufficiently large $I^*>0$, the ensemble of ABPs separates into two subensembles -- one persistently moving to the left [red solid lines in Fig.~\ref{fig4}], and one persistently moving to the right [red dashed lines in Fig.~\ref{fig4}]. See Movie S2 for an exemplary trajectory.
	
	\textit{Parameters: How much inertia is needed?}---The two criteria to observe inverted sedimentation read (in dimensional units): 
	$J\omega \gg \max(\sqrt{Jg_\text{bh}}$,$2\pi\gamma_\text{r})$ and $I^2>2J\omega^2g_\text{bh}$. These expressions reflect that our control scheme requires inertia, but fails if $J$ is too large. How much inertia is needed depends on the system under investigation and the driving frequency. For example, for canonical Janus colloids in a liquid, where inertia is typically neglected, one would need to apply some ac field coupling to the particle orientation (e.g. an ac magnetic or electric field for magnetic or metallodielectric colloids respectively) with a frequency of at least $\omega\sim10^8\text{ Hz}$ to stabilize upward self-propulsion. (Typical parameters $\gamma_\text{r}=8\pi\eta R^3$, $J\sim mR^2\sim\frac{4}{3}\pi\rho R^5$, $\eta\sim10^{-3}\text{ Pa s}$, $R\sim1\text{ }\mu\text{m}$ and $\rho\sim10^3\text{ kg m}^{-3}$ yield $2\pi\gamma_\text{r}/J\sim10^7\text{ Hz}$.)
	In the presence of bottom-heaviness ($g_\text{bh} \sim m_\text{cap}gR/2$~\cite{Wolff2013}) for a cap of mass $m_\text{cap}\sim10^{-15}\text{ kg}$~\cite{Singh2018}, we additionally need a minimum torque of $I\sim10^{-15}\text{ N m}$, which can be achieved e.g. for magnetic colloids in ac magnetic fields~\cite{Han2021}.
	
	Much slower drivings are required when using magnetized vibrated granulates~\cite{ledesma2022magnetized} on tilted plates~\cite{Scholz2018,Scholz2018a} in an ac magnetic field, where inertial effects are substantially stronger. For typical parameters ($m\sim10^{-3} \text{ kg}$, $J\sim10^{-8} \text{ kg m}^2$, $\gamma\sim10^{-2} \text{ kg s}^{-1}$, $\gamma_\text{r}\sim10^{-6} \text{ kg m}^2\text{ s}^{-1}$, $v_0\sim10^{-1} \text{m s}^{-1}$ yield $2\pi\gamma_\text{r}/J\sim10^3\text{ Hz}$), frequencies of $\omega\sim10^4\text{ Hz}$ should be sufficient to observe inverted sedimentation. 
	If the particles are not symmetric but effectively bottom-heavy ($g_\text{bh}\sim10^{-5} \text{ kg m}^2\text{ s}^2$) one additionally needs a torque of at least $I\sim 10^{-2} \text{ N m}$, which can be achieved e.g. based on a spatially uniform ac magnetic field of strength $B\sim$ 10$^{-2}$ T coupling to magnetic dipoles of $m_\text{D} \sim$ 1 A m$^2$~\cite{Cullity2008} embedded into the granular particles. For a plate inclination angle of 3$^{\circ}$, we then obtain $v^*_\text{s}=\frac{mg}{\gamma v_0}\sin($3$^{\circ})\sim0.$5, ${I^*}^2/(2{\omega^*}^2)\sim0.5$ and $g^*_\text{bh}\sim0.1$, so that $\sim 30\%$ of the particles would move upwards [see inset in Fig.~\ref{fig3}].
	
	\textit{Conclusions.}---Our results unveil a generic principle to control the motion of APs by stabilizing fixed points in their orientation dynamics with ac fields. Unlike other schemes for controlling self-propulsion, this scheme does not require an explicit bias, but works even for external fields with a vanishing time average. This offers several advantages: (i) All particles are identically controlled, as opposed e.g. to particles which are placed at different positions in a rotating flow field or in a rotating magnetic field. (ii) There is no need to realize any large-scale gradients as required e.g. for tactic mechanisms. Accordingly, control can be readily switched on and off, which is not easily possible for mechanisms like chemotaxis or thermotaxis, where a complete renewal of the underlying concentration/temperature field would be required. (iii) The mechanism works even for particles with isotropic shape, as opposed e.g. to viscotaxis~\cite{Liebchen2018} or gravitaxis~\cite{TenHagen2014} (and without internal asymmetries in the mass distribution~\cite{Campbell2013}). This scheme could be used in the future e.g. to (dynamically) influence the collective behavior of APs or to segregate mixtures e.g. by inertia (mass) or properties like size or coating geometry, which influence the coupling to the ac field and can lead to purified subensembles at the top wall.

\end{document}